\documentclass[sort&compress,final,5p,times,twocolumn]{elsarticle}

\usepackage{amssymb}
\usepackage{amsmath}
\usepackage{lineno}

\usepackage[hidelinks,colorlinks]{hyperref}

\usepackage{siunitx}

\graphicspath{{images/}}

\setkeys{Gin}{width=0.48\textwidth}

\usepackage[labelsep=period]{caption}
\journal{Nucl. Instr. Meth. Phys. Res. A}

\begin{document}
	

\begin{frontmatter}

	\title{Ultra-Low Noise Mechanically Cooled Germanium Detector}
	
	\author[lbnl]{P. Barton\corref{cor1}}
	\ead{pjbarton@lbl.gov}
	\cortext[cor1]{Corresponding author. Tel.: +1 510 486 6644}
	\author[lbnl]{M. Amman}
	\author[queens]{R. Martin}
	\author[lbnl,ucb]{K. Vetter}
	\address[lbnl]{Lawrence Berkeley National Laboratory, University of California, Berkeley, CA 94720, USA}
	\address[ucb]{Department of Nuclear Engineering, University of California, Berkeley, CA 94720, USA}
	\address[queens]{Queen's University at Kingston, Ontario, Canada}
	
	\begin{abstract}
		
				
		Low capacitance, large volume, high purity germanium (HPGe) radiation detectors have been successfully employed in low-background physics experiments.  However, some physical processes may not be detectable with existing detectors whose energy thresholds are limited by electronic noise.  In this paper, methods are presented which can lower the electronic noise of these detectors.  Through ultra-low vibration mechanical cooling and wire bonding of a CMOS charge sensitive preamplifier to a sub-pF p-type point contact HPGe detector, we demonstrate electronic noise levels below 40\,eV-FWHM.
	\end{abstract}
	
	\begin{keyword}
	Germanium Detector \sep Low Noise \sep Front-End Electronics \sep Mechanical Cooling \sep Reactor Antineutrino \sep Coherent Neutrino Scattering
	\end{keyword}

\end{frontmatter}



\section{Introduction} \label{sec:Introduction}

	High purity germanium (HPGe) detectors offer excellent energy resolution in a relatively large (kg-scale) format with material purity suitable for ultra-low background experiments.  Low capacitance HPGe detectors have demonstrated below 200~eV-FWHM electronic noise~\cite{vetter_recent_2007}, leading to applications in the search for dark matter~\cite{cogent_collaboration_cogent:_2013} and neutrinoless double beta decay~\cite{abgrall_majorana_2014}.  

	The direct detection of antineutrinos is another motivating application.  Nuclear reactor antineutrinos have unique features which make them attractive for nuclear safeguards~\cite{_final_2009}: they cannot be shielded, are a direct result of the fission process, and provide information on the operational power and fissile content of reactor cores.  Conventional antineutrino detectors operate on the principle of inverse beta decay $ \left( \bar{\nu_e} + \text{p} \rightarrow \text{e}^+ + \text{n} \right) $, with nuclear reactor demonstrations employing ton-scale liquid scintillators~\cite{reines_neutrino_1956,bowden_experimental_2007}.
	
	Development of below ton-scale detectors could significantly improve reactor monitoring capabilities.  Consideration has therefore been given to \textit{coherent elastic neutrino-nucleus scattering}, whose enhanced scattering cross section (proportional to the square of the number of atomic neutrons) relative to inverse beta decay provides for several orders of magnitude reduction in detector mass.  In this process, a neutrino is predicted~\cite{freedman_coherent_1974, drukier_principles_1984} to scatter off an atomic nucleus.  The recoiling nucleus then imparts a fraction of its gained energy ($ \sim$20\% for Ge~\cite{barbeau_large-mass_2007}) to the creation of electron-hole pairs~\cite{lindhard_integral_1963} which can then be detected in a suitably low-threshold detector.  Typical reactor antineutrino energies (up to several MeV) would yield ionization signals of only a few hundred electron volts, with a greater number of events at lower energies.  
		
	The majority of germanium recoils would go undetected in a commercially available HPGe detector with an electronic noise of 150~eV-FWHM \cite{barbeau_large-mass_2007}.  Reducing the electronic noise to 50~eV-FWHM would increase the detection rate by two to three orders of magnitude \cite{hogan_clearnu_2012}.  The objective of this work is to lower the energy threshold by reducing the electronic noise of the detector-preamplifier system, specifically by reducing the capacitance and temperature of both the germanium crystal and the front end transistor.
	

\section{Ultra Low Noise Detection System} \label{sec:Methods}
	
	The electronic noise observed at the output of a charge sensitive preamplifier can be described in terms of the equivalent noise charge (ENC)~\cite{spieler_semiconductor_2005}, a sum of voltage, $ 1/f $, and current noise terms:
	\begin{equation} \label{eqn:ENC}
	\begin{aligned}
	\text{ENC}^2 = F_v  \frac{4k_BT}{g_m}  \frac{C_\text{in}^2}{\tau_p} 
	+ F_{1/f} A_f C_\text{in}^2 
	+ F_i \left( 2q_e I_\text{in} \right) \tau_p
	\end{aligned}
	\end{equation}
	where $ F_v $, $ F_{1/f}, $ and $ F_i $ are: voltage, $ 1/f $, and current noise factors defined by the choice of shaping function~\cite{radeka_low-noise_1988}. 
	
	The voltage noise of the FET is proportional to its temperature~$ T $ and inversely proportional to its transconductance~$ g_m $.  The capacitance $ C_\text{in} = C_\text{det} + C_\text{FET} + C_\text{fb} + C_\text{test} + C_\text{stray} $ includes all capacitances at the field effect transistor (FET) input: detector, FET gate, feedback, test, and stray.  Methods for reducing $ C_\text{in} $ include: altering electrode geometries to reduce the detector capacitance ($ C_\text{det} $), selecting a lower input capacitance FET ($ C_\text{FET} $), and reducing the feedback capacitor ($ C_\text{fb} $).  A capacitor for test pulses ($ C_\text{test}) $ at the input should be a very small fraction of $ C_\text{in} $.  The design of crystal and FET support structures should be carefully planned to minimize the stray capacitance ($ C_\text{stray} $) to all nearby conductors.  The voltage noise term is inversely proportional to the peaking time~$ \tau_p $.
	
	The $ 1/f $ noise factor $ A_f $ is dependent on dielectric properties and fabrication processes, however its impact on the ENC is significantly reduced in ultra-low $ C_\text{in} $ systems.  
	
	The current noise $ I_\text{in} $ includes leakage currents from both detector and FET and is scaled by the electron charge $ q_e $.  While not explicitly stated, the leakage current can typically be improved by lowering the temperatures of detector and FET.  The current noise term is directly proportional to $ \tau_p $.  Microphonic noise, not included in \autoref{eqn:ENC}, is generally observed at larger peaking time $ \tau_p $. 

	Several issues with existing HPGe detectors may complicate the further reduction of electronic noise.  Typical cold front end electronics require some form of thermal standoff, complicating their placement very near the detector, thus increasing $ C_\text{in} $.  Liquid cryogens increase operational complexity and induce microphonic noise from boiling.  Extremely low $ C_\text{det} $~(\textless1\,pF) HPGe crystals may be difficult to contact with conventional spring-loaded pins.  Lower temperatures achievable with mechanical cooling improve leakage current and mobility in silicon transistors and germanium detector crystals, but typical cryocoolers introduce excessive microphonic noise.
	
	To resolve these issues, the system employed in this work comprises: an ultra-low vibration mechanically cooled cryostat, housing a p-type point contact HPGe detector, that is wire bonded to a low-capacitance preamplifier-on-a-chip.  The design, specification, and assembly of components are detailed in the remainder of this section.

\subsection{Low Vibration Mechanical Cooling}

	While liquid nitrogen (LN$ _2$) is sufficient for cooling HPGe detectors, the lower temperatures (down to 4\,K) achievable by mechanically refrigerated cryocoolers~\cite{radebaugh_cryocoolers:_2009} are beneficial to leakage currents and carrier mobilities, as the lower phonon population reduces the likelihood of lattice scattering~\cite{aubry-fortuna_electron_2010,morin_lattice-scattering_1954}.  Semiconductor surfaces and electrical contacts made to semiconductors generally have lower leakage currents at lower temperatures~\cite{sze_physics_2006}, thereby reducing noise.  At low enough temperatures, charge carrier concentrations at the FET contacts may be reduced to a level where they are said to ``freeze out''~\cite{selberherr_mos_1989}, degrading or preventing operation.  At extremely low temperatures, charge trapping in the HPGe bulk may be a concern. 
		
\begin{figure}[ht!] 
		\centering
		\includegraphics{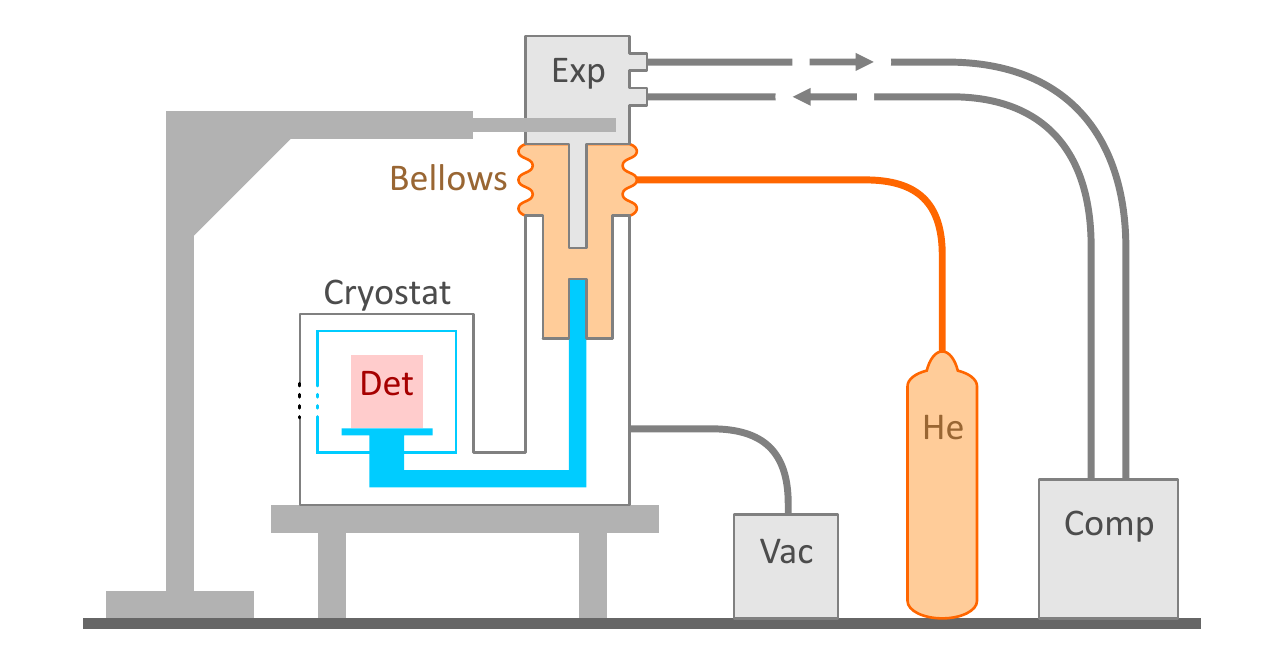}
		\caption{The ultra-low vibration GM cryocooler consists of a compressor and expander (i.e.~cold~head) which cool a separate volume of atmospheric pressure helium, which then cools the cold finger of a vacuum cryostat, housing a point contact HPGe detector and front end electronics. The cryostat and expander are mechanically connected only through a rubber bellows.}
		\label{fig:GM}
	\end{figure}

	Sub-LN$ _2 $ temperatures were achieved in this work with a Gifford-McMahon (GM)~\cite{gifford_gifford-mcmahon_1966} cryocooler (model DE-204) from Advanced Research Systems~\cite{_advanced_2015}.  A 3.5~kW water-cooled compressor was connected through flexible compressed helium lines to a cold head expander affixed to a cantilevered floor stand, while a vacuum cryostat was mounted to a steel table which was isolated from the floor.  Vibrations from the significant displacement of the cold head (up to \SI{100}{\micro m} at {10~m/s$ ^2 $}~\cite{ekin_experimental_2006}) were eliminated with a scheme employed in low temperature optical microscopy (see \autoref{fig:GM}) wherein the cryostat cold finger is disconnected from the from the cold head and heat is instead communicated through atmospheric pressure helium gas.  A flexible rubber bellows contains the helium and a 0.5~psi pressure is maintained during temperature transitions through venting or addition of~99.999\%~pure helium.  This configuration achieved its base temperature of 8\,K in 5~hr, and a Kapton foil heater on the cryostat cold finger enabled operation up to room temperature.
		
	The cryocooler system from ARS was delivered with customizations to their ultra-low vibration optical cryostat (model DMX-20), which was fabricated from nickel-plated aluminum and oxygen-free high-conductivity (OFHC) copper.  The cryostat was then further modified to house our detector and readout electronics. The inner $ \SI{100}{\mm}~\text{diameter by } \SI{100}{mm}~\text{thick} $ chamber was enclosed by an infrared shield held at the primary stage temperature of~$ \SI{\sim40}{K} $.  Small gaps in the shield were included for pumping, but were minimized to reduce the admission of thermal radiation which would increase the observed detector leakage current.  The infrared shield contained an aluminized mylar window, aligned with a 0.33~mm thick beryllium window in the outer vacuum shroud.  The cryostat was actively pumped to $ 10^{-6}$~Torr to reduce the possibility of contaminant adsorption onto the HPGe point contact surface.  Any vibrations from the flexible stainless steel vacuum hose appeared not to have impacted the measured noise performance, as verified by briefly powering down the pump during active measurements.

\subsection{Low Capacitance HPGe Detector}

	The large volume point contact HPGe detector~\cite{luke_low_1989} was originally developed to lower the electronic noise in large mass n-type detectors for the direct detection of weakly interacting particles.  The small electrode of this configuration yields a detector capacitance on the order of 1~pF, compared to the tens of picofarads for traditional coaxial germanium detectors.  The similar p-type point contact (PPC) detector has found utility in several neutrino and astroparticle physics experiments~\cite{barbeau_large-mass_2007, abgrall_majorana_2015}. 
		
	The PPC HPGe detector used in this work was employed previously to demonstrate the low-noise capabilities of the low-mass front end (LMFE) electronics~\cite{barton_low-noise_2011} developed for the M\textsc{ajorana} D\textsc{emonstrator}~\cite{abgrall_majorana_2015}.  This 20~mm~diameter by 10~mm~thick detector originally had a 1.5~mm diameter point contact with a concave dimple for alignment of a tensioned pin contact.  The detector previously had a capacitance of 0.47~pF and had exhibited sub-pA leakage current through many temperature and vacuum cycles over several years.  The outer n-type hole-blocking contact was formed by lithium diffusion, while the bipolar blocking point contact was formed by sputtered amorphous silicon (a-Si)~\cite{amman_amorphous-semiconductor-contact_2007}.
		
	The detector described above was modified to obtain a lower capacitance through the combination of an even smaller point contact electrode and the use of wire bonding for the interconnection between the detector and the front end electronics.  To modify the detector, the point contact face was hand-lapped (600~grit SiC) to remove the dimple which was incompatible with wire bonding. A 4:1~HF:HNO$ _3 $ etch was performed to remove any lapping damage. A new layer of a-Si was sputtered onto the point contact face, and an \SI{8}{k\angstrom} aluminum film was evaporated through a 0.75~mm diameter hole shadow mask to form the point contact electrode.  The crystal was mounted in a spring-loaded, indium-lined aluminum clamp, held at a positive high voltage bias.  This assembly was mounted onto a boron nitride (Saint Gobain AX05) high voltage insulator.  A silicon diode temperature sensor was affixed to a prototype crystal to ensure the detector temperature adequately tracked that of the cold finger.  
	
	The 0.75~mm point contact of the detector was ultrasonically wedge bonded with 2~mil aluminum wire to a trace on the PCB shared by the wire bond to the preamplifier input (see \autoref{fig:CUBE-PPC}).   

\subsection{CMOS Front End Electronics}

	Electronic noise in the charge sensitive preamplifier is typically dominated by contributions from the first field effect transistor (FET) in the front end electronics~\cite{spieler_semiconductor_2005}, for which two specific devices are commonly employed.  The junction FET (JFET) has been the traditional choice~\cite{goulding_design_1969,nashashibi_low_1990} due to its low $ 1/f $ noise and low voltage noise when cooled.  Similar in operation is the metal oxide semiconductor FET (MOSFET) as found in complementary MOSFET (CMOS) processes, which provide for integration of the entire charge sensitive preamplifier as well as additional signal processing.
	
	The optimization of noise in the typical silicon JFET requires elevation of the temperature (above $ \sim $120~K) with some thermal standoff or heater, which can increase stray capacitance at the JFET gate.  Custom JFETs which function down to 4\,K~\cite{nawrocki_silicon_1988} are not available with the low (\textless\,1\,pF) capacitance required for ultra-low noise detector front end electronics.   While the transconductances of both JFET and MOSFET improve continually as the temperature is lowered~\cite{szelag_transconductance_1997}, the relatively high minimum operating temperature of JFETs directly support the use of MOSFETs below LN$ _2 $ temperatures where HPGe leakage currents may also improve.  
	
	Although MOSFETs tend to have higher $ 1/f $ noise factors than their JFET counterparts~\cite{horowitz_art_2015}, an ultra-low capacitance system will reduce the impact of the $ 1/f $ noise term in the ENC (see \autoref{eqn:ENC}).  Low MOSFET capacitance can be obtained through proper design~\cite{de_geronimo_mosfet_2005,bertuccio_progress_2007}, while low JFET capacitance is typically limited to the selection from available devices.  
		
\begin{figure}[ht!]
		\centering
		\includegraphics{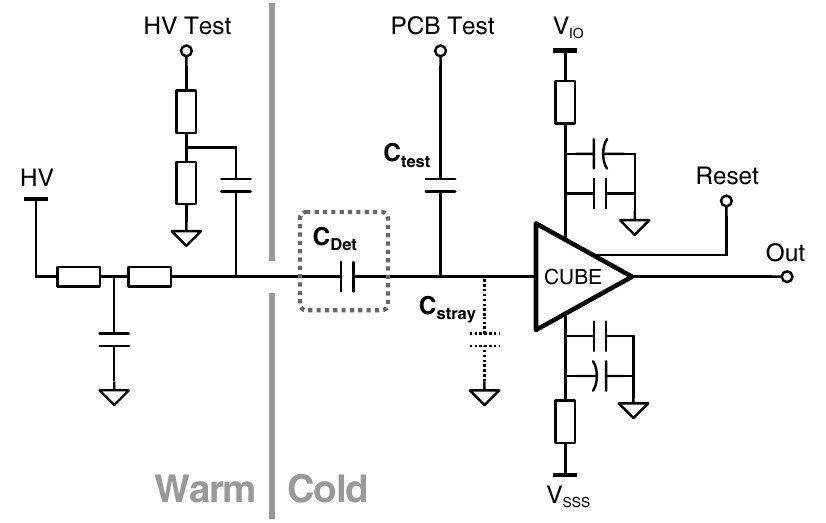}
		\caption{Front end signal chain, including CUBE power supply filters.  Test pulse capacitors are included both before (HV Test) and after (PCB Test) the detector.}
		\label{fig:schematic}
	\end{figure}

	While CMOS front ends have been integrated with higher capacitance HPGe detectors~\cite{pullia_cryogenic_2010,riboldi_low_2007}, even lower capacitance CMOS front end electronics have been developed for low capacitance ($ \sim $0.1~pF) silicon drift detectors (SDD)~\cite{bombelli_high_2012,bertuccio_silicon_2015}.  The ``CUBE'' application specific integrated circuit (ASIC), developed by XGLab~\cite{_xglab_2015} is such a CMOS charge sensitive pulse-reset preamplifier-on-a-chip, exhibiting remarkably low electronic noise~\cite{bombelli_low-noise_2010,bombelli_cube_2011}.  Encouraged by applications of the CUBE to other HPGe detectors~\cite{krings_high-resolution_2014,krings_multi-element_2015,tartoni_monolithic_2015}, we integrated this ASIC with a LBNL ultra-low capacitance PPC detector (\autoref{fig:schematic}).

	\begin{figure}[ht!]
		\centering
		\includegraphics[width=\columnwidth]{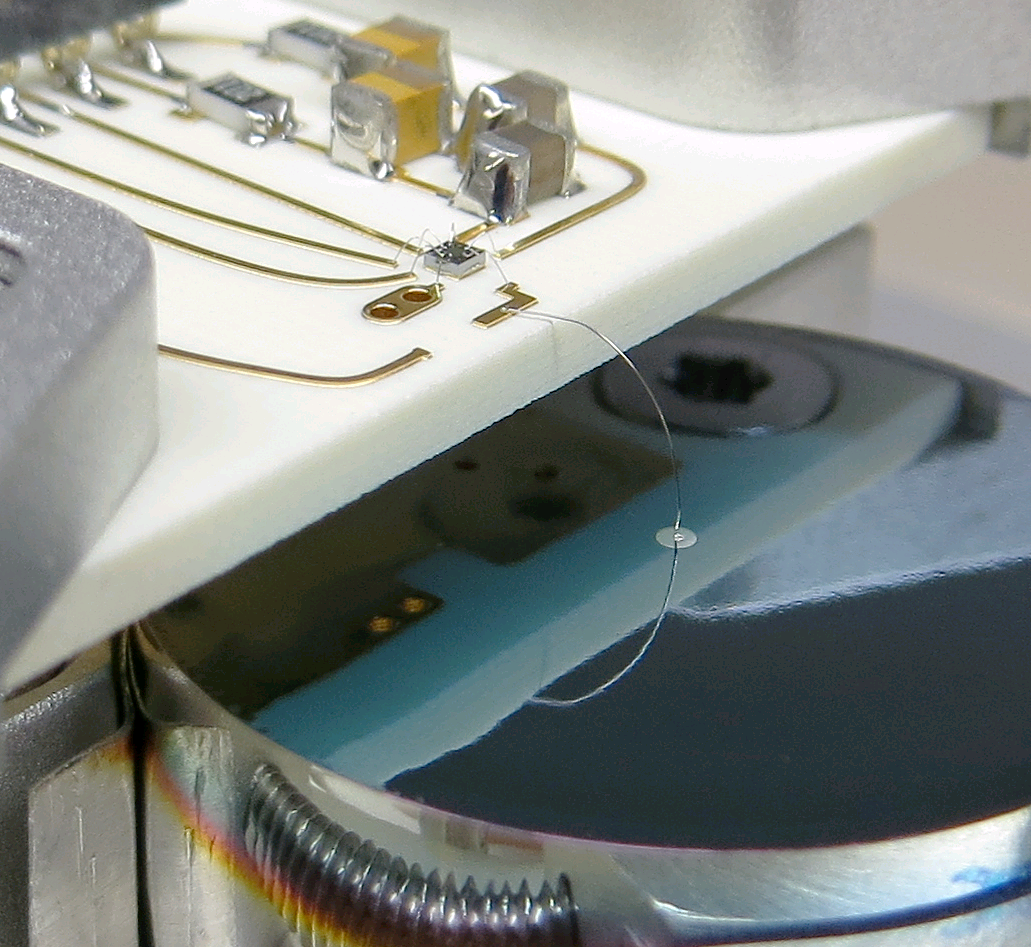}
		\caption{PPC detector and CUBE ASIC wire bonded to low dielectric loss PCB with power supply filters.}
		\label{fig:CUBE-PPC}
	\end{figure}

	The $ 0.75 \times \SI{0.75}{\mm^2}$ CUBE die (prototype version PRE\_024) was epoxied to a floating pad on a 60~mil Rogers-4350 low dielectric loss printed circuit board (PCB) with electroless nickel immersion gold (ENIG) traces (see \autoref{fig:CUBE-PPC}).  Caution was exercised in preventing MOSFET damage from electrostatic discharge.  Pads on the ASIC were ultrasonically wedge bonded to PCB traces with 1~mil Al(1\% Si) wire.  The board also contained two identical RC bypass filters (see \autoref{fig:schematic}) for the ${ V_\text{SSS}=~{-3.0\text{ V }} \text{ and } V_\text{IO}=~{+3.3\text{ V }}}$ ASIC power supplies, composed of size 0805 components: 
	$ \SI{200}{\ohm} \text{ metal film}, \SI{47}{\micro F} \text{ tantalum} + \SI{47}{\nano F} \text{ ceramic C0G} $.  A separate test pulse trace was provided on the PCB, whose parasitic capacitance to the input pad was simulated and measured to be 0.010 pF.

	Signal and power wires from the CUBE PCB were routed through several infrared and vacuum feedthroughs to an external XGLab-supplied ``Bias Board'' (ver.~7), which filtered the $ {\pm10 \text{ VDC }}$ from a standard bench power supply, supplied reset logic and level control, and buffered the CUBE output with a gain of $ -2.25 $.

	Detector bias was supplied by a Canberra 3002D high voltage power supply.  A high voltage RC low-pass filter (100~M$ \Omega $, 10~nF) removed significant voltage fluctuations, and an associated resistive divider provided for injection of charge through a step voltage pulse onto the detector capacitance.

\section{Results} \label{sec:Results}

	The two parameters of a HPGe detector with the greatest influence on electronic noise are its depleted capacitance and its leakage current. In this section, we first present the detector capacitance and leakage current measurements made with our ultra-low noise detector system. Measurements are then given for the energy resolution and equivalent noise charge as a function of temperature and peaking time.

\subsection{Detector Capacitance}

	The capacitance of the detector was measured by applying a known voltage step $ \Delta V_{\text{pulse}} $ onto the high voltage electrode of the detector (see \autoref{fig:schematic}).  The resulting charge pulse measured by the preamplifier, as calibrated with the 59.5~keV peak from an Am-241 source, is then equal to the product of $ \Delta V_{\text{pulse}} $ and the depleted capacitance $ C_\text{det} $.  As the high voltage bias is increased, the depleted region grows, and the capacitance decreases until the detector is fully depleted.

	\begin{figure}[ht!]
		\centering
		\includegraphics{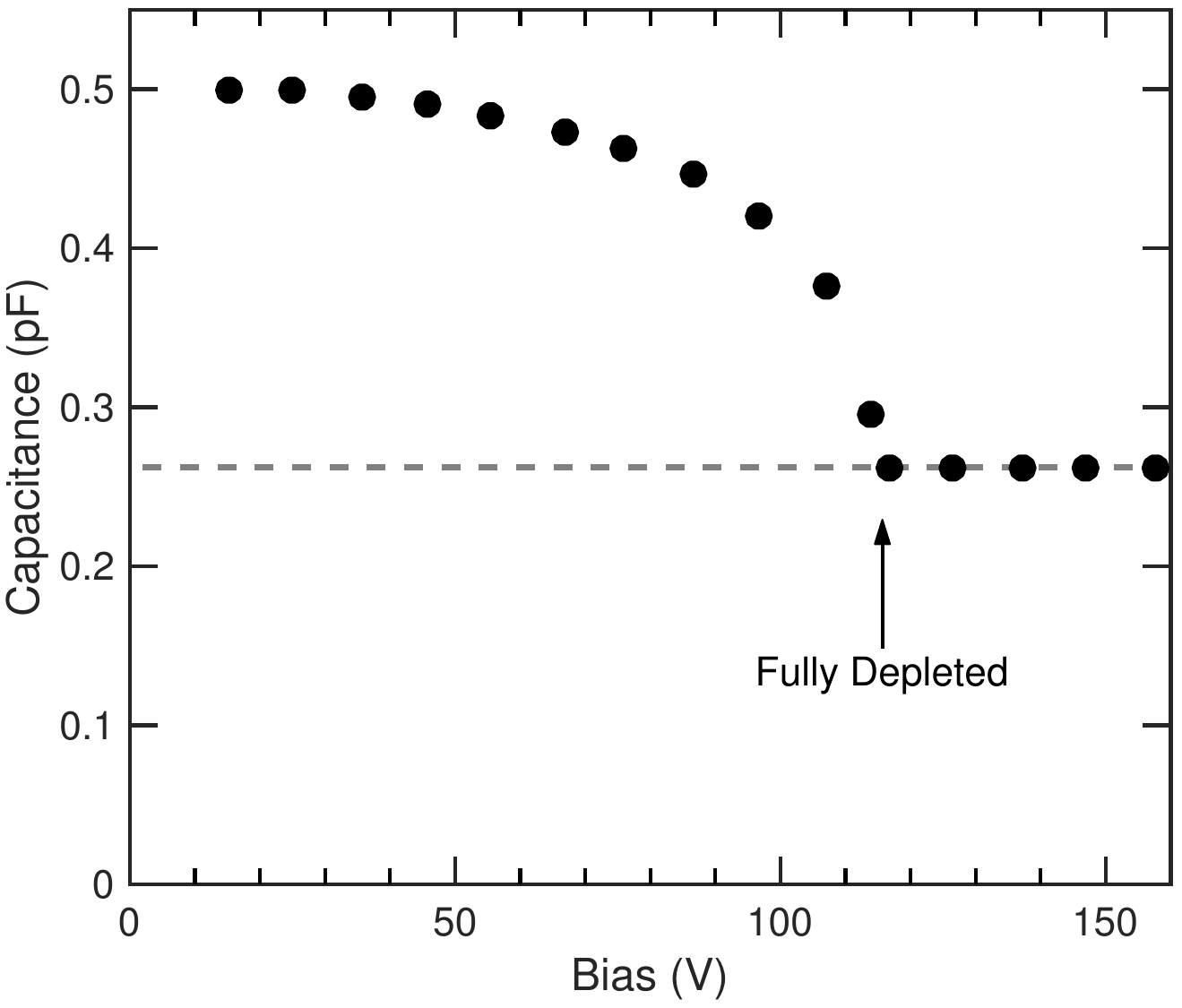}
		\caption{PPC detector capacitance as measured by pulse injection through a high voltage filter capacitor.  The full depletion voltage is 120 V, beyond which which the detector capacitance is 0.26~pF.}
		\label{fig:CV}
	\end{figure}
	
	The full depletion voltage of the crystal was determined to be 120\,V from measurements of the capacitance as a function of applied bias (\autoref{fig:CV}).  The 0.26~pF capacitance at full depletion of this detector (a 45\%~reduction from the previous detector design) agreed well with 3D electrostatic simulations performed with the finite element modeling package COMSOL~\cite{_comsol_2015}.

\subsection{Leakage Current}
	
	The leakage current ${ I_\text{leak} = Q_\text{FS}/t_\text{reset} }$ was determined from $ Q_\text{FS} $, the calibrated full scale charge and $ t_\text{reset} $, the time between preamplifier reset pulses.  The leakage current as a function of applied bias was measured at multiple temperatures as the cryostat was cooled.  At 500\,V above full depletion, the current remained less than twice that at full depletion, indicating stable performance of the detector when overdepleted.  

	\begin{figure}[ht!]
		\centering
		\includegraphics{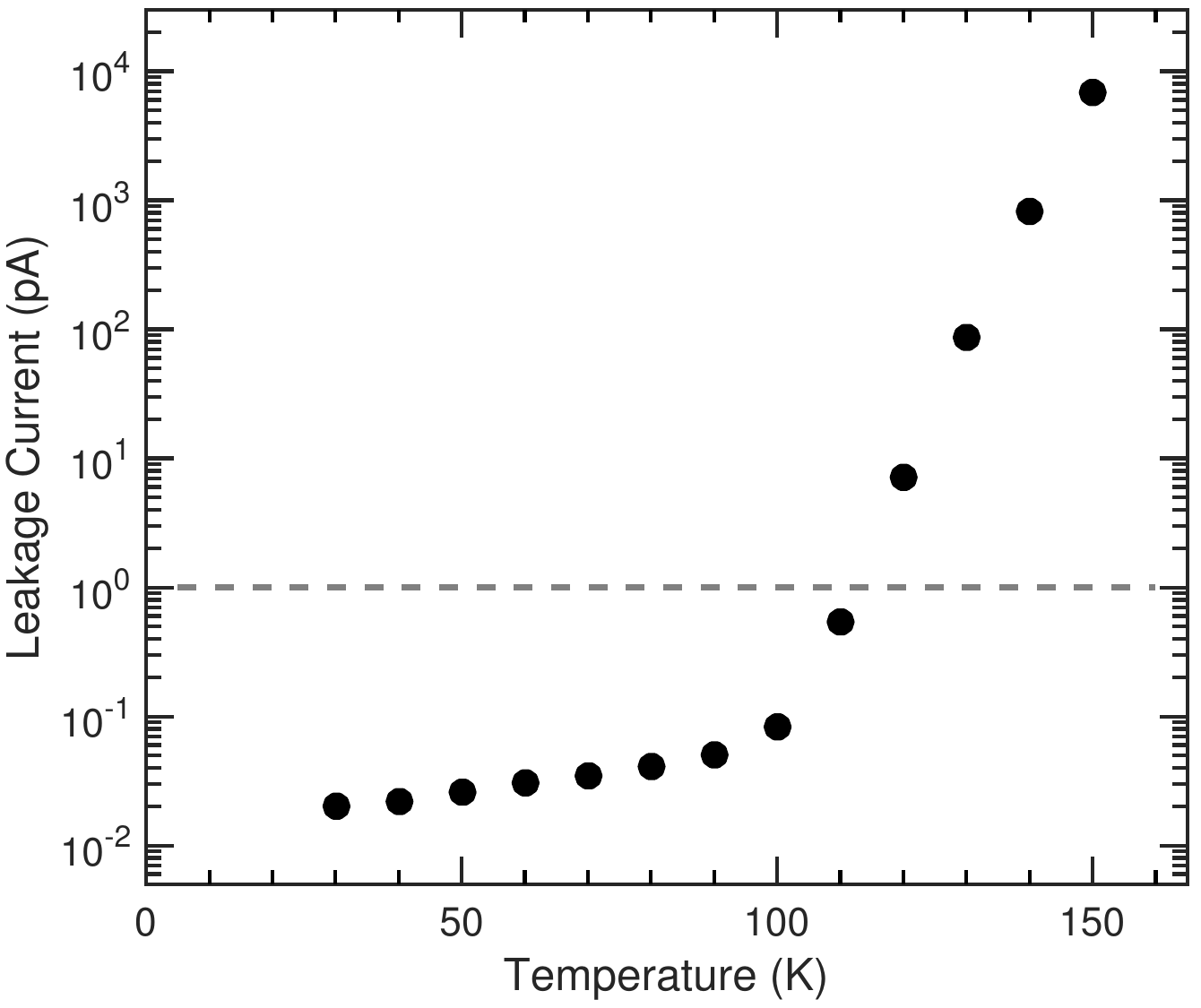}
		\caption{Leakage current as measured by pulse reset intervals, when the detector was biased at \SI{150}{V}.  The dashed line at 1~pA is for visual reference.}
		 \label{fig:leakage}
	\end{figure}

	The leakage currents measured at 150\,V were observed (\autoref{fig:leakage}) to drop by approximately an order of magnitude with every 10\,K of cooling from 150\,K to 100\,K.  Below 100\,K, the leakage current was \textless\,0.1~pA, with a minimum of 0.020~pA at 30\,K.  Additional tests will be needed to determine the specific origin of the currents measured at low temperatures.
	
\subsection{Energy Resolution}

\begin{figure}[ht!] 
		\centering
		\includegraphics{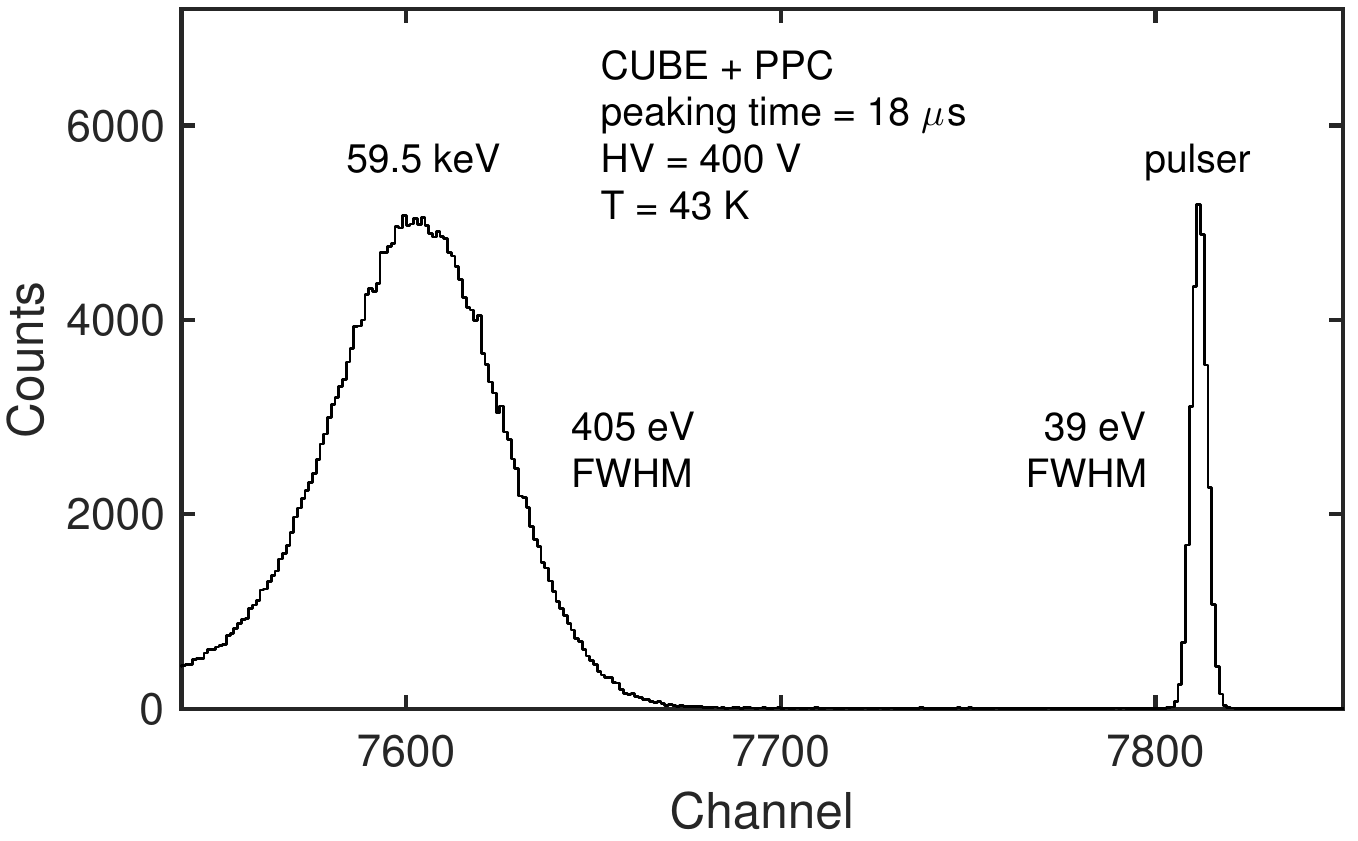}
		\caption{Spectrum of $ ^{241} $Am indicating the 59.5~keV gamma line from the source and a pulser peak with a width of 39~eV-FWHM.}
		\label{fig:39eV}
	\end{figure}

	The flux from an uncollimated $ ^{241} $Am source was directed through the beryllium window toward both the lithium contact and point contact faces of the detector.  The buffered preamplifier output was filtered by a pair of Canberra 2026x semi-gaussian shaping amplifiers with peaking times that ranged from \SI{0.2}{\micro\second}~to~\SI{53}{\micro\second}.  Pulse heights were measured by an Amptek MCA8000D multi-channel analyzer (MCA).  The resulting energy spectrum at 43\,K is shown in  \autoref{fig:39eV} for a 400\,V bias and \SI{18}{\micro s} peaking time.  The electronic pulser peak width of 39~eV-FWHM is among the lowest measured with a HPGe detector of this size. 
		
	Calibrated baseline rms voltages from the shaping amplifier were sampled at 50~MSa/s on an Agilent DSOX-3054A 500~MHz bandwidth oscilloscope to calculate the ENC in electrons-rms.  The standard deviation (i.e. rms voltage) of several million samples between radiation events compared closely to rms voltages from an analog 20~MHz bandwidth rms meter (Boonton 93A).	The 39~eV-FWHM value was independently verified with the 5.6 electrons-rms determined by the calibrated 2.6~mV-rms baseline voltage of the shaping amplifier.  

	\begin{figure}[ht!] 
		\centering
		\includegraphics{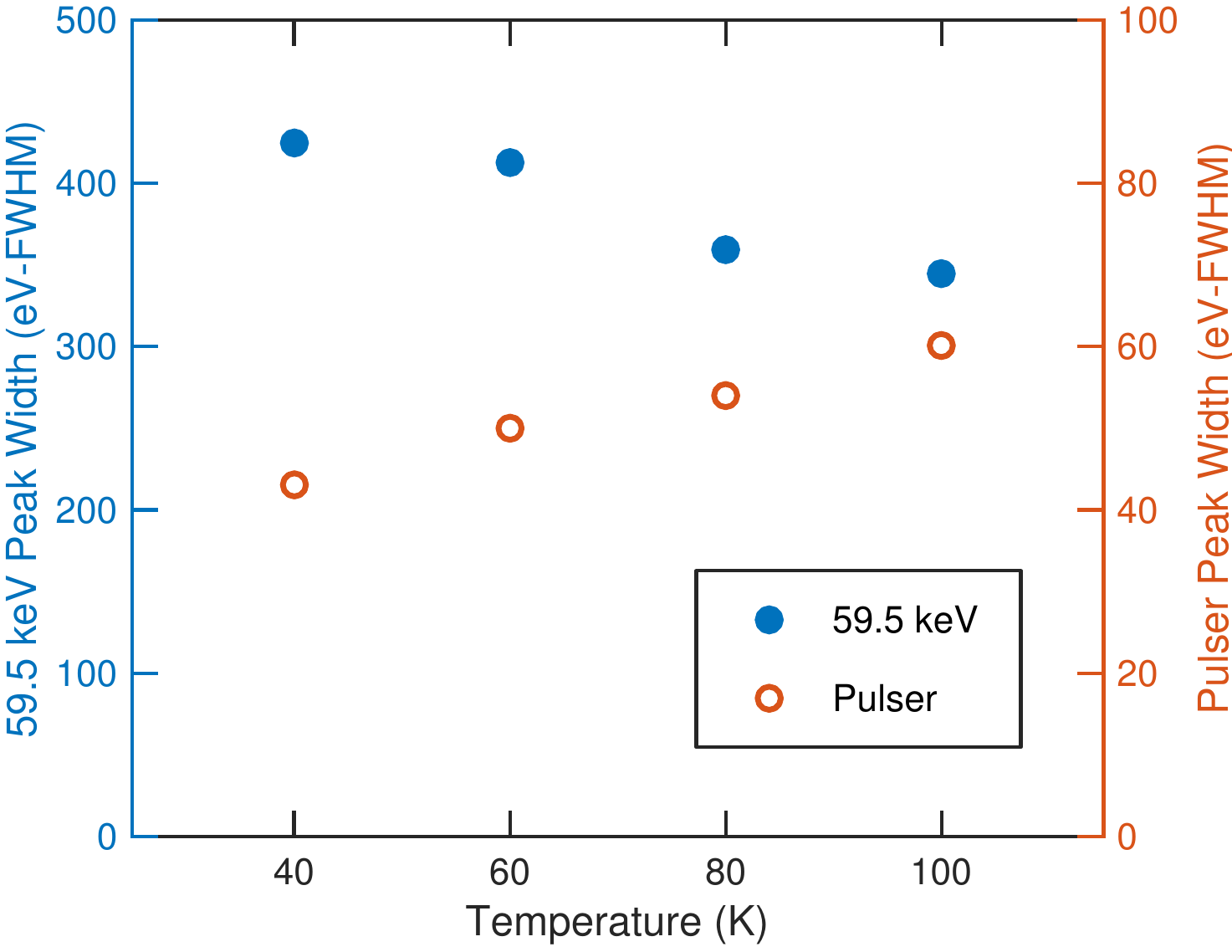}
		\caption{Spectrum peak widths for $ ^{241} $Am 59.5~keV (left axis) and pulser (right axis), at an applied bias of \SI{150}{V}, and peaking time of \SI{53}{\micro\second}.}
		\label{fig:eVT}
	\end{figure}
	
	Gaussian fits to the 59.5~keV peak appeared to be broader than that expected when assuming a reasonable Fano factor.  Preliminary measurements indicated an improvement in resolution as the temperature was raised above 40\,K (see  \autoref{fig:eVT}), while a slight degradation was observed in the electronic noise.  The minimum resolution observed at 59.5~keV was 345~eV-FWHM, which represents a derived Fano factor of 0.122, neglecting incomplete charge collection.  No signal processing was performed to remove slow signals from degraded surface events.  Results will be reported separately on the impact of temperature and bias on the charge collection and contact properties of this particular detector and the associated noise of the preamplifier ASIC.  The full characterization of the low energy spectral performance of this crystal was left for future work with more suitable low energy contacts.

\subsection{Electronic Noise}
	
\begin{figure}[ht!] 
		\centering
		\includegraphics{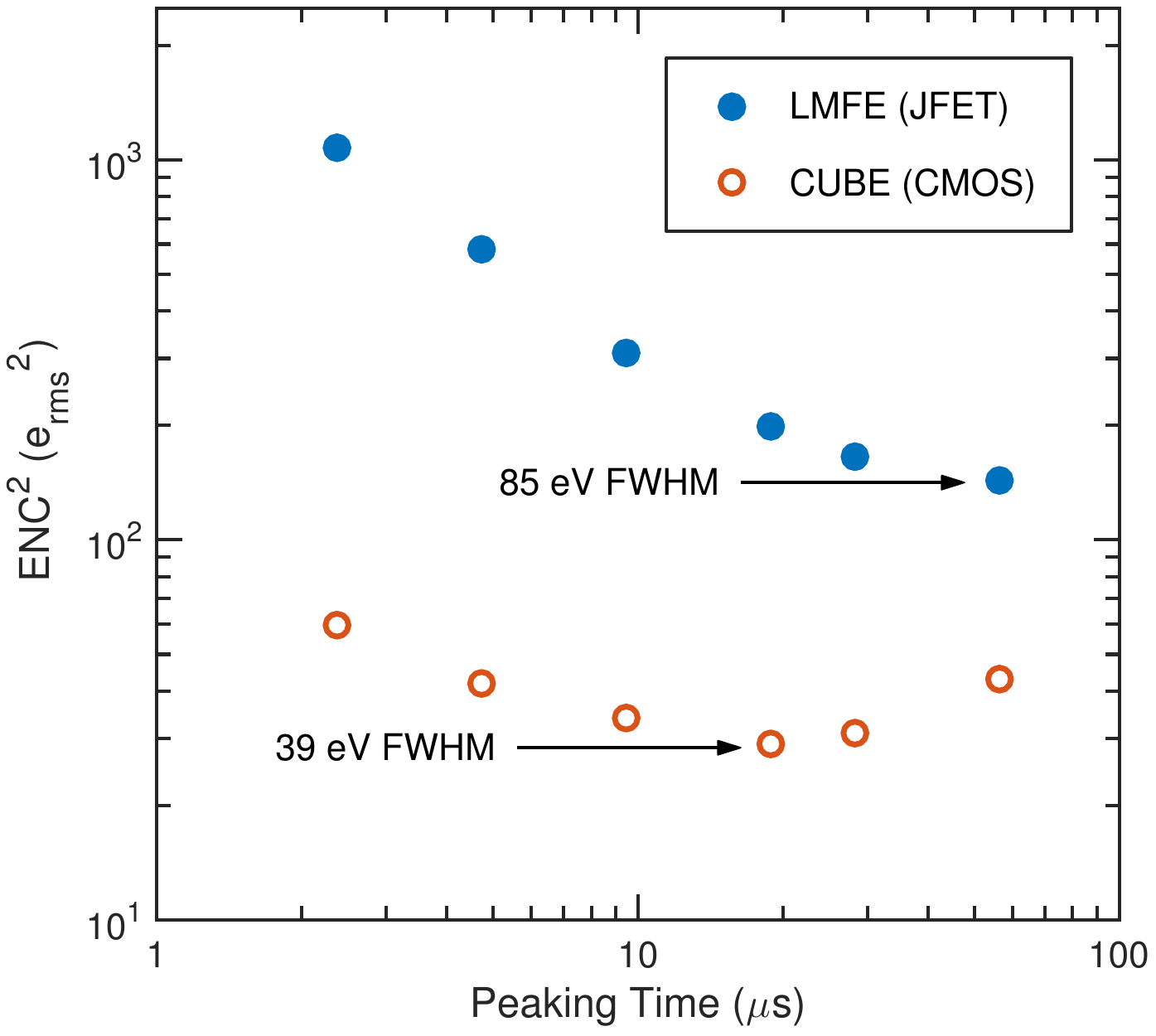}
		\caption{Equivalent noise charge (from spectral peak widths) versus peaking time of PPC with CUBE CMOS ASIC at 43\,K and 400\,V bias.  The minimum resolution of 39~eV-FWHM from pulser peak widths is equivalent to a noise of 5.6~electrons-rms.  The previous low noise limit of 85~eV-FWHM is shown for this crystal in its higher capacitance configuration with LN$ _2 $ cooling and a cold JFET~\cite{barton_low-noise_2011}.}
		\label{fig:ENC}
	\end{figure}

	The ENC curve (vs.~peaking time) of the low capacitance PPC with CMOS preamplifier in the low vibration cryostat at 43\,K (\autoref{fig:ENC}) indicates the minimum pulser peak width of 39~eV-FWHM.  The previous lowest noise performance of this crystal with a low mass JFET front end was 85~eV-FWHM as indicated in the ENC curve from Ref.~\cite{barton_low-noise_2011}.  The primary factor responsible for this improvement was the decreased capacitance of detector and FET, each about 40\% to 50\% of their previous values.  This lower capacitance improved both voltage noise and the relative contribution of $ 1/f $ noise.    

\begin{figure}[ht!] 
		\centering
		\includegraphics{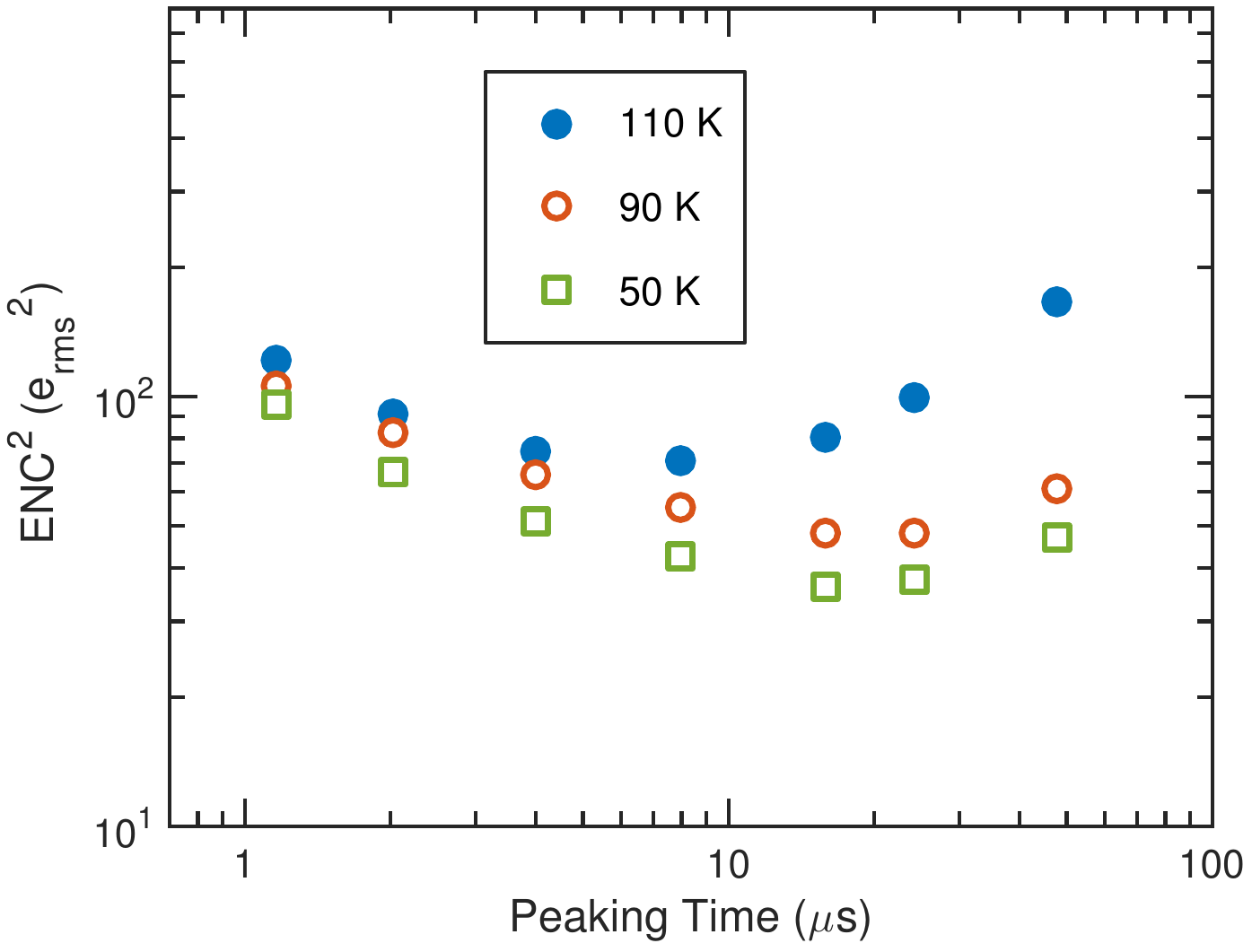}
		\caption{Equivalent noise charge as in \autoref{fig:ENC}, illustrating minimum resolutions of 42~eV-FWHM (50\,K), 49~eV-FWHM (90\,K) and 60~eV-FWHM (110\,K).  The significant current noise component at 110\,K is consistent with the 0.5~pA leakage current from \autoref{fig:leakage}.}
		\label{fig:ENCT}
	\end{figure}

	The lower temperature improved leakage currents in both FET and detector, and reduced thermally related voltage noise at shorter peaking times.  Continuous reduction of voltage and current noise was observed (\autoref{fig:ENCT}) as the temperature was lowered from \SI{110}{K} to 50\,K (and below).

\section{Discussion}

	Three specific barriers to lowering electronic noise in detection systems were identified and overcome in this work to create a uniquely low noise HPGe detector technology.

	First, excessive vibrations from conventional mechanical cooling (30 to 80\,K) would significantly degrade the performance of low noise germanium detectors.  Boiling liquid cryogens also cause microphonics which limit low noise performance at longer peaking times.  Our detector system operated down to 30 K and eliminated microphonics by employing a technique from cryogenic microscopy which employs an atmospheric pressure heat exchange gas with a conventional GM cryocooler.
	
	Second, the detector capacitance of commercially available HPGe detectors is typically limited by point contact sizes and stray capacitances (e.g. from spring-loaded pin contacts).  We reduced the detector capacitance to 0.26~pF by wire bonding to a 0.75~mm diameter detector electrode.
	
	Third, the noise of the JFET in conventional detectors must be optimized by raising its current and temperature, increasing the complexity of operation.  We integrated an ultra-low capacitance PPC detector with a commercially available ultra-low noise CMOS preamplifier-on-a-chip ASIC, whose performance improves down to 30 K.

\subsection{Implications for Antineutrino Detection}

\begin{figure}[ht!] 
		\centering
		\includegraphics{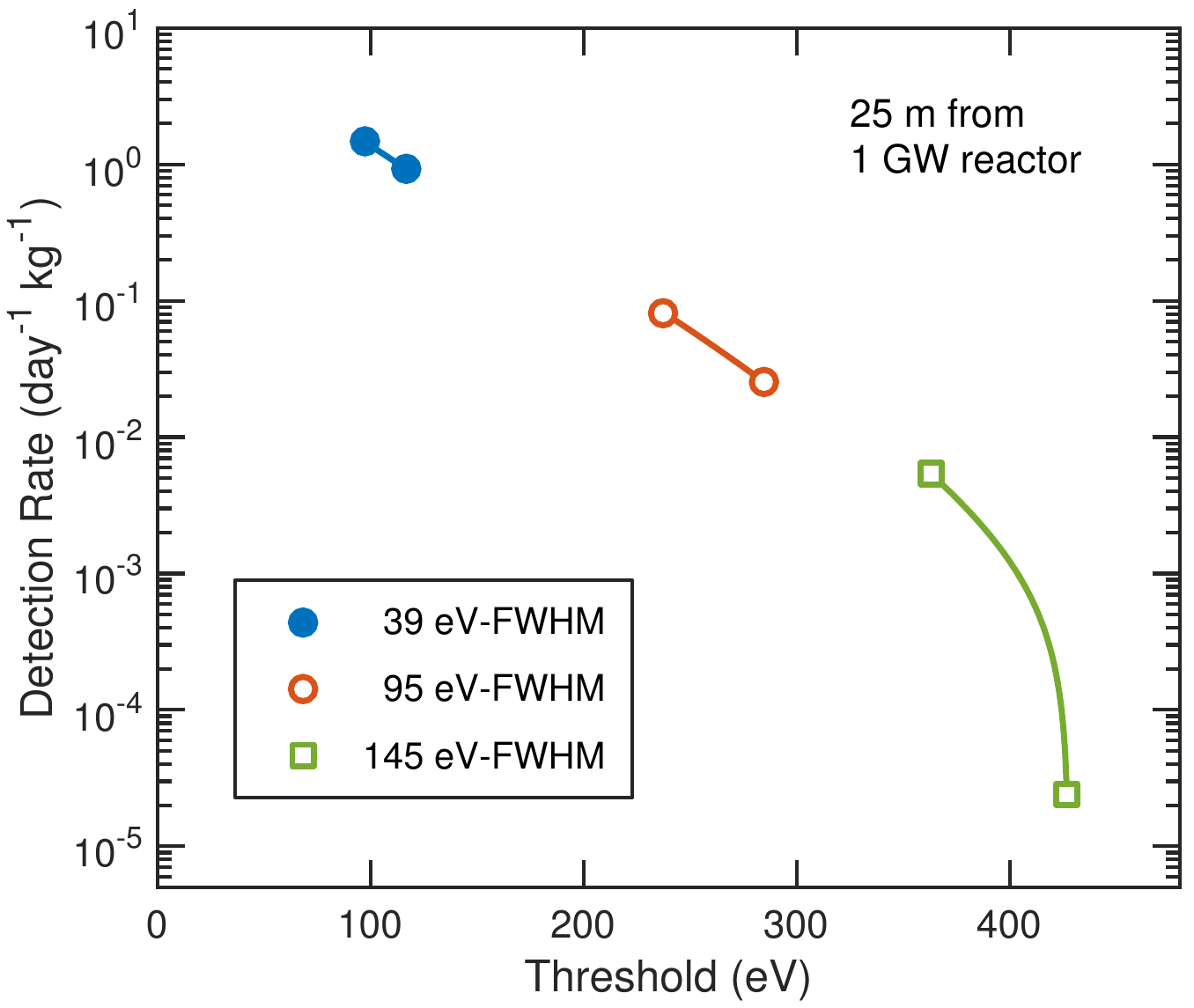}
		\caption{Simulated number of reactor antineutrinos detected, per day, per kg of Ge, at 25~m from a 1~GW reactor.  The minimum threshold ranges (2.5~to~3.0~times FWHM noise) are shown for three electronic noise levels.}
		\label{fig:n_per_kgday}
	\end{figure}

	Three key elements of a successful coherent elastic neutrino-nucleus scattering measurement are: low radioactive backgrounds, a large detector mass, and a low energy threshold.  To illustrate the impact of electronic noise and energy threshold on the antineutrino detection rate, we consider a typical energy distribution of electron antineutrinos $ \bar{\nu}_e $ from a 1~GW nuclear power reactor.  The expected number of antineutrinos detected~\cite{hogan_clearnu_2012} through Ge nucleus recoils in a given time with a given detector mass is illustrated (\autoref{fig:n_per_kgday}) for three electronic noise levels.  
	
	The noise level of 95~eV-FWHM represents the lowest noise performance of the low mass front end demonstrated during development of the M\textsc{ajorana} D\textsc{emonstrator}~\cite{barton_low-noise_2011}.  In this work, a detector technology was presented with a 39~eV-FWHM noise level which would increase the anticipated $ \bar{\nu}_e $ detection rate by several orders of magnitude from the previous deployment at 145~eV-FWHM~\cite{barbeau_large-mass_2007}, while maintaining capabilities for larger mass detectors in a suitably low background configuration.

\subsection{Future Applications}

	Despite the modest size of the detector employed in this work, previous experience indicates that the same low capacitance and low leakage current can be reliably obtained from a kg-scale crystal at a higher bias.  Adaptations of this technology may prove beneficial for the scaling of experiments such as the M\textsc{ajorana} D\textsc{emonstrator}.  The GM cryocooler with heat exchange gas could allow for greater flexibility in the design and implementation of heavy shielding.  Both compressors and expanders can be located well outside of the sensitive region of the experiment.  The integrated nature of the full preamplifier ASIC reduces the number of vacuum feedthroughs required for multi-detector experiments.
	
	One area deserving specific attention for low background experiments is the ASIC power supply filter network which must remain close to the ASIC.  Bypass filter capacitors of sufficient capacitance typically contain ceramics which may be too radioactive for low background experiments.  Progress has been made in the development of parylene-based resistors and capacitors~\cite{dhar_low-background_2015}, and could be extended to multi-layer large-value capacitors.
	
	Another potential application of the ultra-low noise germanium recoil detector described herein is the direct detection of intermediate energy neutrons (10~keV to 1~MeV) where conventional liquid scintillators lose efficiency.  New detection modalities may be realized for the inspection of nuclear processes involving intermediate energy neutrons, perhaps in combination with a broad energy range of gamma rays. 
	
	The lower energy threshold limit of charge-based ionization detection in HPGe has yet to be reached.  Ongoing development of charge sensitive preamplifiers have reduced electronic noise levels to as low as 1.5~electrons-rms~\cite{bertuccio_silicon_2015}.  Similar efforts to reduce detector and stray capacitances could lead to another factor of two improvement in the electronic noise of HPGe detectors.
 	
\section{Conclusions} \label{sec:Conclusions}

	The low energy threshold of low capacitance HPGe detectors enables their use as germanium recoil detectors for weakly interacting particles (e.g. neutrinos and neutrons).  While HPGe detector noise has been improved in recent years by optimizing JFET-based approaches with liquid cryogens, we explored the electronic noise benefits of lowering temperatures of both detector and front-end electronics, and integrating state of the art low capacitance X-ray detector electronics with lower capacitance point contact HPGe detectors.
	
	The ultra-low noise germanium test stand developed herein presents a unique and important opportunity for the study of fundamental physics of high purity germanium detectors and low noise electronics over a broad range of temperatures.  Charge carrier mobility and leakage currents both improve at lower temperatures.  Wire bonding allows for a more point contact-like detector capacitance.  Vibrations from mechanical cryocoolers are eliminated with an atmospheric pressure heat exchange gas.  These efforts made possible an ultra-low vibration, mechanically cooled, CMOS ASIC preamplifier-on-a-chip, wire bonded to a PPC HPGe detector, with 39~eV-FWHM electronic noise at 43\,K.  

\section*{Acknowledgments}
	
	We thank Luca Bombelli of XGLab for his support in adapting the CUBE ASIC.  The UCB-led Nuclear Science and Security Consortium (NSSC) was instrumental in obtaining instrumentation from Advanced Research Systems, to whom we also extend our thanks for helpful discussions. Daniel Hogan provided useful calculations of antineutrino rates.  We thank Rhonda Witharm for her wirebonding expertise.  Paul Luke provided useful discussions throughout the project.
	
	This work was performed under the auspices of the U.S. Department of Energy by Lawrence Berkeley National Laboratory under Contract DE-AC02-05CH11231.  This project was funded by the US Department of Energy, National Nuclear Security Administration, Office of Defense Nuclear Nonproliferation Research and Development (DNN R\&D).  Support was also provided by the Department of Energy, National Nuclear Security Administration under Award Number: DE-NA0000979 through the Nuclear Science and Security Consortium.

%

\section*{}

	\bibliographystyle{elsarticle-num} 
	\bibliography{LowNoiseGe}
	
\end{document}